\documentclass[twocolumn,showpacs]{revtex4}

\usepackage{amsmath, amssymb, graphicx, url}

\newcommand{\ud}{\mathrm{d}}
\renewcommand{\vr}{{\bf r}}
\newcommand{\vq}{{\bf q}}
\newcommand{\vA}{{\bf A}}

\newcommand{\ve}{{\bf e}}
\newcommand{\vxi}{{\mbox{\boldmath$\xi$}}}
\newcommand{\Lag}{\mathcal{L}}
\newcommand{\erf}{\mathrm{erf}}
\newcommand{\T}{{\mathrm{T}}}

\begin{document}

\title{Scattering of quantum wave packets by shallow potential
  islands:\\ A quantum lens}

\author{Arseni Goussev$^{1,2}$ and Klaus Richter$^3$}

\affiliation{$^1$Department of Mathematics and Information Sciences,
  Northumbria University, Newcastle Upon Tyne, NE1 8ST, United
  Kingdom\\ $^2$Max Planck Institute for the Physics of Complex
  Systems, N{\"o}thnitzer Stra{\ss}e 38, D-01187 Dresden,
  Germany\\ $^3$Institute for Theoretical Physics, University of
  Regensburg, D-93040 Regensburg, Germany}

\date{\today}

\begin{abstract}
  We consider the problem of quantum scattering of a localized wave
  packet by a weak Gaussian potential in two spatial dimensions. We
  show that, under certain conditions, this problem bears close
  analogy with that of focusing (or defocusing) of light rays by a
  thin optical lens: Quantum interference between straight paths
  yields the same lens equation as for refracted rays in classical
  optics.

\end{abstract}


\maketitle

\section{Introduction}

The intrinsic connection between the motion of classical particles on
one hand and the propagation of quantum matter waves on the other has
occupied minds of scientists since the early days of quantum
theory. De Broglie was one of the first to realize that ``for both
matter and radiations $\ldots$ it is necessary to introduce the
corpuscle concept and the wave concept at the same time''
\cite{deB29wave}. Subsequently, invaluable contributions of Ehrenfest,
Van Vleck, Feynman, Gutzwiller, Maslov, among many others, shaped our
current understanding of quantum wave propagation in terms of
interference of classical trajectories. However, a number of important
questions concerning quantum-classical correspondence remain
open. These questions fall under the scope of the area of mathematical
physics known as {\it quantum chaos} \cite{Gut90Chaos,Sto99Quantum,
  Haa10Quantum}.

The motion of a classical particle can be conveniently described by
means of a phase-space trajectory. The Heisenberg's uncertainty
principle however does not allow for the notion of the classical
trajectory to be directly carried over to quantum theory: the
particle's position and momentum can not be specified simultaneously.

One natural extension of the classical concept of a point in the phase
space is provided by a localized quantum wave packet that can be
parametrized by its mean position and momentum, and dispersion
quantifying the phase-space extent of the wave packet. According to
the Ehrenfest theorem \cite{Tan07Introduction}, the time evolution of
the mean position and momentum is governed, for short enough times, by
the classical equations of motion. In other words, the wave packet
center follows the corresponding classical trajectory.

An issue of the wave packet spreading, i.e., how the dispersion
depends on time, is however much more complex. Loosely speaking, there
are two main mechanisms of the spreading: (i) a classical-like
broadening of the wave packet due to forces exerted by an external
potential, and (ii) an intrinsically quantum-mechanical spreading
dictated by the uncertainty principle. Due to the interference nature
of quantum dynamics, the overall spreading is not a simple ``sum'' of
the two contributions, but rather a more intricate process.

A natural question arises: is there an {\it intuitive} and, at the
same time, {\it quantitative} theoretical description of the
phenomenon of quantum spreading? In this paper, we develop such a
description, based on the short-wavelength approximation to quantum
dynamics, for the simple system of a two-dimensional quantum wave
packet scattered by a weak Gaussian potential. In particular, we show
that the quantum scattering process bears a close mathematical analogy
with the phenomenon of focusing (or defocusing) of light rays by a
thin lens, and can be described using the ``language'' of geometrical
optics. Interestingly, on the quantum side the use of the Eikonal
approximation \cite{Sakurai94}, i.e. including interference of {\em
  straight} paths, yields the same thin lens equation as derived in
classical optics from {\em refracted} light rays.

Our theoretical approach provides an intuitive picture of the wave
packet spreading \cite{Vandegrift04,Heller06}, and its quantitative
predictions are found in good agreement with results of an ``exact''
numerical solution of the time-dependent Schr{\"o}dinger equation.

\section{Theory}

\begin{figure}[ht!]
\includegraphics[width=3.3in]{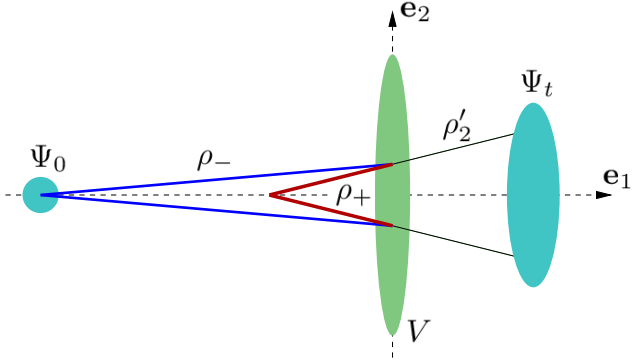}
\caption{(Color online) Schematic illustration of the scattering
  system under consideration.}
\label{fig1}
\end{figure}

We consider a quantum particle of mass $m$ that evolves in the
two-dimensional position space under the influence of the external
potential
\begin{equation}
  V(\vq) = V_0 \, e^{-\vq \cdot \vA \vq} \,.
\label{eq:potential}
\end{equation}
Here, $V_0$ quantifies the strength of the potential, $\vq$ is a
column vector representing the particle's position, and $\vA$ is a
2-by-2 orthogonal matrix with eigenvalues $a_1$ and $a_2$
corresponding, respectively, to orthonormal (column) eigenvectors
$\ve_1$ and $\ve_2$. In other words,
\begin{equation}
  \vA = (\ve_1 \; \ve_2) \; \mathrm{diag}(a_1, a_2) \; (\ve_1 \; \ve_2)^\T \,,
\label{eq:matrix_A}
\end{equation}
where $|\ve_1| = |\ve_2| = 1$ and $\ve_1 \cdot \ve_2 =
0$. Hereinafter, the dot ``$\cdot$'' stands for the scalar product,
and the superscript ``$^{\mathrm{T}}$'' denotes the matrix
transposition. Equation~(\ref{eq:potential}) describes a Gaussian
potential ``island'' centered at $\vq={\bf 0}$, see
Fig.~\ref{fig1}. The spatial extent of the island is characterized by
the length $l_1 = 1/\sqrt{a_1}$ in the direction of the vector $\ve_1$
and by $l_2 = 1/\sqrt{a_2}$ in the direction of $\ve_2$.

The dynamics of the quantum particle is fully described by its
time-dependent wave function, $\Psi_t (\vq)$. The latter is related to
the initial wave function, $\Psi_0(\vq)$, by means of a propagator $K$
in accordance with
\begin{equation}
  \Psi_t(\vq) = \int \ud \vq' \, K(\vq,\vq';t) \Psi_0(\vq') \,.
\label{eq:prop-1}
\end{equation}
Here, the $\vq'$-integration runs over the whole two-dimensional
plane. In general, the propagator $K$ is determined by solving the
time-dependent Schr{\"o}dinger equation, 
\begin{equation}
  \left( i \hbar \frac{\partial}{\partial t} + \frac{\hbar^2}{2m}
  \nabla_{\vq}^2 - V(\vq) \right) K(\vq,\vq';t) = 0 \,,
\label{eq:Schroedinger}
\end{equation}
subject to appropriate boundary and initial conditions, e.g., see
Ref.~\cite{Bar89Elements}. Generally, this task is analytically
formidable, and one is left to resort to numerical computations. In
certain cases however analytical progress is made possible by
constructing ``sensible'' approximations to the propagator. In this
paper, we focus on one such approximation commonly referred to as {\it
  semiclassics}.

The semiclassical (or short-wavelength) approximation of the
propagator $K(\vq,\vq';t)$ is formulated in terms of {\it classical}
trajectories $\gamma$ that start at the point $\vq'$ at time $0$ and
end at $\vq$ at time $t$. More precisely, $\gamma = \left\{
\vr(\tau)\,, \; \tau \in [0,t] \right\}$, such that $m \frac{\ud^2
  \vr}{\ud \tau^2} + \nabla V(\vr) = {\bf 0}$ with $\vr(0) = \vq'$ and
$\vr(t) = \vq$. The approximate propagator, commonly referred to as
the Van Vleck-Gutzwiller propagator, can be written as
\cite{Gut90Chaos}
\begin{equation}
  K(\vq,\vq',t) = \frac{\sqrt{D_{\gamma}}}{2\pi i\hbar} \exp \left(
  \frac{i}{\hbar} S_{\gamma} - \frac{i\pi}{2} \nu_{\gamma} \right) \,,
\label{eq:prop-2}
\end{equation}
where 
\begin{equation} \label{eq.S-1}
  S_\gamma \equiv S [\vr] (t) = \int\limits_0^t \ud \tau \left\{
  \frac{m}{2} \left( \frac{\ud \vr}{\ud \tau}(\tau) \right)^2 - V\big(
  \vr(\tau) \big) \right\}
\end{equation}
is Hamilton's principle function along the trajectory $\gamma$,
\begin{equation} \label{eq.stab}
D_{\gamma} = |\det (-\nabla_{\vq'} \nabla_{\vq} S_{\gamma})| 
\end{equation}
is the
stability factor of $\gamma$, and $\nu_{\gamma}$ is the so-called
Maslov index, counting the number of conjugate points along $\gamma$;
as far as our problem is concerned, the Maslov index is identically
zero, $\nu_{\gamma} = 0$.

The approximation (\ref{eq:prop-2}) is known to be reliable when
applied to wave functions $\Psi_0 (\vq)$ representing a quantum
particle of sufficiently high kinetic energy $E_0$. We additionally
assume that the Gaussian potential, given by Eq.~(\ref{eq:potential}),
is weak compared to the particle's kinetic energy, $|V_0| \ll
E_0$. The validity of this condition is at the heart of our ``shallow
potential island'' approximation corresponding to the Eikonal
approximation in scattering theory \cite{Sakurai94}. In this case (see
Appendix~\ref{app1}), $S_\gamma$ can be well approximated by
$S_{\gamma_0}$, where $\gamma_0$ is the straight, free-particle
trajectory leading from $\vq'$ to $\vq$ in time $t$, i.e., $\gamma_0 =
\left\{ \vq \tau/t + \vq' (1-\tau/t)\,, \; \tau \in [0,t]
\right\}$. Thus, we write
\begin{equation}
  S_\gamma \simeq S_{\gamma_0} = \frac{m}{2t} |\vq-\vq'|^2 - \int_0^t
  \ud \tau \, V\left( \vq'+\frac{\tau}{t}(\vq-\vq') \right) \,.
\label{eq:S-2}
\end{equation}
The integral in the right-hand side of Eq.~(\ref{eq:S-2}) can be
straightforwardly evaluated to equal
\begin{align}
   &\frac{\sqrt{\pi} V_0 t }{2 \mathcal{A}} \exp \left( -\frac{(\vq
    \cdot \vA \vq)(\vq' \cdot \vA \vq')-(\vq \cdot \vA
    \vq')^2}{\mathcal{A}^2} \right) \nonumber\\ &\times \left[
    \erf\left( \frac{\vq \cdot \vA (\vq-\vq')}{\mathcal{A}} \right) -
    \erf\left( \frac{\vq' \cdot \vA (\vq-\vq')}{\mathcal{A}} \right)
    \right] \label{eq:integral}
\end{align}
with
\begin{equation}
  \mathcal{A} \equiv \sqrt{(\vq-\vq') \cdot \vA (\vq-\vq')} \,.
\label{eq:A_cal}
\end{equation}
Expression~(\ref{eq:integral}), and therefore Eq.~(\ref{eq:S-2}), can
be further simplified by taking into account the identity
\begin{equation}
  (\vq \cdot \vA \vq)(\vq' \cdot \vA \vq')-(\vq \cdot \vA \vq')^2 = |
  \vq \times \vq' |^2 \det\vA \,,
\label{eq:vec_identity}
\end{equation}
where ``$\times$'' denotes the vector product. This yields
\begin{align}
  &S_\gamma \simeq \frac{m}{2t} |\vq-\vq'|^2 - \frac{\sqrt{\pi} V_0 t
  }{2 \mathcal{A}} \exp \left( -\frac{| \vq \times \vq' |^2
    \det\vA}{\mathcal{A}^2} \right) \nonumber\\ &\times \left[
    \erf\left( \frac{\vq \cdot \vA (\vq-\vq')}{\mathcal{A}} \right) -
    \erf\left( \frac{\vq' \cdot \vA (\vq-\vq')}{\mathcal{A}} \right)
    \right] \,. \label{eq:S-3}
\end{align}

A substitution of Eq.~(\ref{eq:S-3}) into Eq.~(\ref{eq:prop-2}) leads
to an explicit, closed-form expression for the semiclassical
propagator, $K(\vq,\vq',t)$, and, therefore, provides the complete
solution of the time-dependent scattering problem in the
short-wavelength regime.

The expression for the propagator, $K(\vq,\vq',t)$, becomes especially
simple and allows for an intuitive interpretation in the following
special case. Let us consider a setup, in which the ``receiver'' $\vq$
and the ``source'' $\vq'$ lie on the opposite sides of and at almost
the same distance, large compared to $l_1$, from the center of the
Gaussian scattering potential, and in which the vector $(\vq-\vq')$ is
nearly aligned with one of the principal directions (taken, for
concreteness, to be $\ve_1$) of the potential island. In other words,
we are interested in the asymptotic form of the function
$K(\vq,\vq',t)$ in the case that
\begin{equation}
  \vq = L \ve_1 + \vxi \,, \quad \vq' = -L \ve_1 + \vxi'
\label{eq:L-def}
\end{equation}
and
\begin{equation}
  |\vxi|, |\vxi'|, l_1 \ll L \,.
\label{eq:special}
\end{equation}

Substituting Eq.~(\ref{eq:L-def}) into Eqs.~(\ref{eq:A_cal}) and
(\ref{eq:S-3}), taking into account that $\erf(\pm z) \rightarrow \pm
1$ as $z \rightarrow +\infty$, and keeping only terms to the leading
order in $|\vxi|/L$ and $|\vxi'|/L$ in the argument of the exponential
function, we obtain
\begin{align}
  S_\gamma \simeq &\frac{m}{2 t} | 2 L \ve_1 + \vxi-\vxi' |^2
  \nonumber\\ &- \frac{\sqrt{\pi}}{2} \frac{l_1}{L} V_0 t \exp \left(
  -\frac{| \ve_1 \times (\vxi + \vxi') |^2}{(2 l_2)^2} \right)
  \,. \label{eq:S-4}
\end{align}
Then, using the basis representations $\vxi = \xi_1 \ve_1 + \xi_2
\ve_2$ and $\vxi' = \xi'_1 \ve_1 + \xi'_2 \ve_2$ in
Eq.~(\ref{eq:S-4}), and further assuming that
\begin{equation}
  \xi_2, \xi'_2 \ll l_2 \,,
\label{eq:special-2}
\end{equation}
we write, approximately,
\begin{equation}
  S_\gamma \simeq - \frac{\sqrt{\pi}}{2} \frac{l_1}{L} V_0 t +
  S_{\gamma}^{(1)} + S_{\gamma}^{(2)} \,,
\label{eq:S-5}
\end{equation}
where
\begin{equation}
   S_{\gamma}^{(1)} = \frac{m}{2 t} (2 L + \xi_1-\xi'_1)^2
\label{eq:S1}
\end{equation}
and
\begin{equation}
   S_{\gamma}^{(2)} = \frac{m}{2 t} (\xi_2-\xi'_2)^2 +
   \frac{\sqrt{\pi}}{8} \frac{l_1}{L (l_2)^2} V_0 t \,
   (\xi_2+\xi'_2)^2 \,.
\label{eq:S2}
\end{equation}
In view of Eqs.~(\ref{eq:S-5}--\ref{eq:S2}), the stability factor (\ref{eq.stab})
along the trajectory $\gamma$ can be written as $D_\gamma \simeq
D_\gamma^{(1)} D_\gamma^{(2)}$, with $D_\gamma^{(1)} =
|-\partial_{\xi_1} \partial_{\xi'_1} S_\gamma^{(1)} | = m/t$ and
$D_\gamma^{(2)} = |-\partial_{\xi_2} \partial_{\xi'_2} S_\gamma^{(2)}
| = m/t - \sqrt{\pi} l_1 V_0 t / (4 L (l_2)^2)$, provided that
\begin{equation}
  t^2 \ll \frac{m (l_2)^2}{|V_0|} \frac{L}{l_1} \,.
\label{eq:t-limit}
\end{equation}
In fact, Eq.~(\ref{eq:t-limit}) states a necessary condition for
Eq.~(\ref{eq:S2}) to constitute a ``healthy'' perturbative expansion
of $S_\gamma^{(2)}$ in powers of $V_0$.

The full semiclassical propagator, Eq.~(\ref{eq:prop-2}), can now be
written as
\begin{align}
  K (\vq,\vq',&t) \simeq \exp\left(-i \frac{\sqrt{\pi}}{2}
  \frac{l_1}{L} \frac{V_0 t}{\hbar} \right) \nonumber\\ &\times K_0
  (L+\xi_1, -L+\xi'_1 ; t) \, K_V (\xi_2,\xi'_2;t)
  \,, \label{eq:prop-3}
\end{align}
where
\begin{equation}
  K_0 (z,z',\tau) \equiv \sqrt{\frac{m}{2 \pi i \hbar\tau}} \exp
  \left( i \frac{m}{2 \hbar \tau} (z-z')^2 \right)
\label{eq:K0}
\end{equation}
is the free-particle propagator describing the motion of the particle
in the $\ve_1$-direction, while
\begin{align}
  K_V (z,z',\tau) \equiv &K_0 (z,z',\tau) \sqrt{1 -
    \frac{\sqrt{\pi}}{4} \frac{l_1}{L} \frac{V_0 \tau^2}{m (l_2)^2}}
  \nonumber\\ &\times \exp \left( i \frac{\sqrt{\pi}}{8} \frac{l_1}{L}
  \frac{V_0 \tau}{\hbar} \frac{ (z+z')^2}{(l_2)^2}
  \right) \label{eq:KV}
\end{align}
accounts for the wave function spreading in the orthogonal,
$\ve_2$-direction. Clearly, $K_V \rightarrow K_0$ as $V_0 \rightarrow
0$, recovering the free-particle limit.

We now observe that
\begin{align}
  &K_V (z,z',\tau) = \int_{-\infty}^{+\infty} \!\! \ud \zeta \,
  K_0(z,\zeta,\tau/2) \nonumber\\ &\times \exp \left[ i
    \frac{\sqrt{\pi}}{2} \frac{l_1}{L} \frac{V_0 \tau}{\hbar} \left( 1
    - \frac{\sqrt{\pi}}{4} \frac{l_1}{L} \frac{V_0 \tau^2}{m (l_2)^2}
    \right)^{-1} \frac{\zeta^2}{(l_2)^2} \right] \nonumber\\ &\times
  K_0 (\zeta,z',\tau/2) \,. \label{eq:KV-representation}
\end{align}
Equation~(\ref{eq:KV-representation}) is an identity, and can be
verified straightforwardly by evaluating the Gaussian integral in the
right-hand side. Then, after Eq.~(\ref{eq:t-limit}) is taken into
account, Eq.~(\ref{eq:KV-representation}) reduces to
\begin{align}
  K_V &(z,z',\tau) \simeq \int_{-\infty}^{+\infty} \!\! \ud \zeta \,
  K_0(z,\zeta,\tau/2) \nonumber\\ &\times \exp \left[ i
    \frac{\sqrt{\pi}}{2} \frac{l_1}{L} \frac{V_0 \tau}{\hbar}
    \frac{\zeta^2}{(l_2)^2} \right] K_0 (\zeta,z',\tau/2)
  \,. \label{eq:KV-representation-2}
\end{align}

The physical picture offered by Eq.~(\ref{eq:KV-representation-2}) is
as follows. The propagator $K_V$, evolving a quantum state during time
$\tau$, can be view as a result of three consecutive operations: (i) a
free-particle propagation during time $\tau/2$, (ii) an instantaneous
phase change, or ``kick'', of the quantum state, and (iii) another
free-particle propagation during $\tau/2$. This interpretation, and
the physical meaning of the kick operator, becomes apparent when the
evolving quantum state is given by a Gaussian wave packet. To this
end, we consider as initial state the two-dimensional wave packet
\begin{equation}
  \Psi_0(\xi'_1, \xi'_2) \equiv \psi_0^{(1)}(\xi'_1; \rho_1) \,
  \psi_0^{(2)}(\xi'_2; \rho_2)
\label{eq:WP_0}
\end{equation}
with
\begin{align}
  \psi_0^{(1)}(z; \rho) &\equiv \left( \frac{1}{\pi \sigma^2}
  \right)^{\frac{1}{4}} \exp \left[ i \frac{m v}{\hbar} \left(
    \frac{z^2}{2 \rho} + z \right) \right]
  \,, \label{eq:wp0-1}\\ \psi_0^{(2)}(z; \rho) &\equiv \left(
  \frac{1}{\pi \sigma^2} \right)^{\frac{1}{4}} \exp \left( i \frac{m
    v}{\hbar} \frac{z^2}{2 \rho} \right) \,. \label{eq:wp0-2}
\end{align}
Here, $\rho_1$ and $\rho_2$ (usually termed radii of curvature
\cite{GD05Lyapunov}) are two, generally complex-valued,
parameters. The real-valued function $\sigma = \sigma(\rho)$, defined
in accordance with
\begin{equation}
  \frac{1}{\sigma^2} \equiv \frac{m v}{\hbar} \, \Im \left(
  \frac{1}{\rho} \right) = - \frac{m v}{\hbar} \, \frac{\Im
    \rho}{|\rho|^2} \,,
\label{eq:sigma}
\end{equation}
quantifies the position-space dispersion of the wave
packet. Furthermore, $v$ specifies the average velocity (and $m v$ the
average momentum) of the particle.

Now, acting with the propagator (\ref{eq:prop-3}) on the initial state
(\ref{eq:WP_0}), which is assumed to be spatially localized around the
position vector $\vq'=-L \ve_1$ (or around the origin in the
$\vxi'$-coordinate frame), we obtain the quantum state after time $t$
locally, in the vicinity of the point $\vq = L \ve_1$ (or around the
origin in the $\vxi$-coordinate frame):
\begin{equation}
  \Psi_t(\xi_1, \xi_2) \simeq e^{-i \frac{\sqrt{\pi}}{2} \frac{l_1}{L}
    \frac{V_0 t}{\hbar}} \, \psi_t^{(1)}(\xi_1; \rho_1) \,
  \psi_t^{(2)}(\xi_2; \rho_2) \,,
\label{eq:WP_t}
\end{equation}
where
\begin{align}
  \psi_t^{(1)}(z; \rho) &\equiv \int_{-\infty}^{+\infty} \!\! \ud
  \zeta \, K_0(L+z, -L+\zeta, t) \, \psi_0^{(1)}(\zeta; \rho)
  \,, \label{eq:wpt-1}\\ \psi_t^{(2)}(z; \rho) &\equiv
  \int_{-\infty}^{+\infty} \!\! \ud \zeta \, K_V(z,\zeta,t) \,
  \psi_0^{(2)}(\zeta; \rho) \,. \label{eq:wpt-2}
\end{align}
As we are concerned with the semiclassical limit, it is reasonable to
expect that, in the course of its time evolution, the quantum wave
packet remains concentrated around the corresponding classical
trajectory. This means that the center of the wave packet, starting
from the point $-L \ve_1$ at time 0, reaches the point $L \ve_1$ in
time $t$, such that
\begin{equation}
  v t = 2 L \,.
\label{eq:time}
\end{equation}
It is at this instant that $\Psi_t (\xi_1,\xi_2)$ is localized around
the origin in the $\vxi$-coordinate frame, and that the propagator
approximation, given by Eq.~(\ref{eq:prop-3}), proves the most
useful. 

Fixing the time $t$ in accordance with Eq.~(\ref{eq:time}) and
evaluating the Gaussian integral in Eq.~(\ref{eq:wpt-1}), we obtain
\begin{equation}
  \psi_t^{(1)}(z; \rho_1) = e^{i \phi_1} \, \psi_0^{(1)}(z; \rho_1 + v
  t) \,,
\label{eq:wpt-1new}
\end{equation}
where $\phi_1 = E_0 t / \hbar - (1/2) \arg (1 + v t / \rho_1)$, and
\begin{equation}
  E_0 \equiv \frac{m v^2}{2} \,,
\label{eq:E_cl}
\end{equation}
denoting the kinetic energy of the corresponding classical
particle. The physical interpretation of Eq.~(\ref{eq:wpt-1new}) is
that the $\ve_1$-component of the wave packet retains its Gaussian
shape, as its center travels in space on top of the corresponding
classical trajectory (and in agreement with the Ehrenfest
theorem). The spreading of the $\ve_1$-component of the wave packet is
entirely described by the linear transformation of the corresponding
radius of curvature, $\rho_1 \rightarrow \rho_1 + v t$, and, in the
weak potential limit, this spreading is not affected by the external
potential.

We now focus on the time evolution of the $\ve_2$-component of the
wave packet. As before, we keep the time $t$ fixed in accordance with
Eq.~(\ref{eq:time}). Substituting Eq.~(\ref{eq:KV-representation-2})
into Eq.~(\ref{eq:wpt-2}), and successively evaluating two Gaussian
integrals, we obtain
\begin{equation}
  \psi_t^{(2)}(z; \rho_2) = e^{i \phi_2} \, \psi_0^{(2)}(z; \rho'_2) \,,
\label{eq:wpt-2new}
\end{equation}
where $\rho'_2 = \rho_+ + v t / 2$,
\begin{equation}
  \frac{1}{\rho_+} = \frac{1}{\rho_-} + \frac{1}{f} \,,
\label{eq:lens}
\end{equation}
$\rho_- = \rho_2 + v t / 2$,
\begin{equation}
  f \equiv \frac{1}{\sqrt{\pi}} \frac{E_0}{V_0} \frac{(l_2)^2}{l_1}
  \,,
\label{eq:focal_length}
\end{equation}
and $\phi_2 = -(1/2) \big[ \arg (\rho'_2 / \rho_+) + \arg (\rho_- /
  \rho_2) \big]$.

The physical picture of the wave packet spreading, offered by the central 
Eqs.~(\ref{eq:wpt-2new}--\ref{eq:focal_length}), bears close analogy
with the focusing (or defocusing) of light rays by a thin optical
lens. Indeed, the well-known thin lens equation, $[\mathrm{object \;
    distance}]^{-1} + [\mathrm{image \; distance}]^{-1} =
[\mathrm{focal \; length}]^{-1}$, can be readily recovered from
Eq.~(\ref{eq:lens}) by interpreting $-\rho_-$ and $\rho_+$ as the
``distances'' from, respectively, the object and its image to a lens
of the focal length $f$; the role of the lens is played here by the
Gaussian potential island, see Fig.~\ref{fig1}.

It is important to point out that the above analogy between the wave
packet scattering in quantum mechanics and the ray focusing in optics
is not a trivial one: in the quantum-mechanical case, the
``distances'' $-\rho_-$ and $\rho_+$ are intrinsically complex-valued
and can only be related to the true distances, encountered in optics,
in a nonlinear way.

\medskip

Finally, we note that our simple wave-packet-propagation construction
can be generalized to arbitrary times $t$. This generalization can be
summarized as follows. Suppose that the initial wave packet is
centered around a point with coordinates $(Q,0)$, where $Q<0$, and
that the initial wave function is given by the product $\psi_0^{(1)}
(q_1-Q; \rho_1) \, \psi_0^{(2)} (q_2; \rho_2)$. Then, at a later time
$t$, the wave function is given, up to an overall phase factor, by
$\psi_0^{(1)} (q_1-Q-vt; \rho_1+vt) \, \psi_0^{(2)} (q_2; \rho'_2)$,
where
\begin{equation}
  \rho'_2 = \left\{
  \begin{array}{ll}
    \rho_2 + v t \;, \quad & t < |Q|/v \\ \rho_+ + v (t - |Q|/v) \;,
    \quad & t \geq |Q|/v
  \end{array} \right.
\label{eq:summary}
\end{equation}
and $\rho_+$ is determined from Eq.~(\ref{eq:lens}) with $\rho_- =
\rho_2 + |Q|$.

\section{Numerical confirmation}

We now confirm the validity of the ``quantum lens'' formulae, given by
Eqs.~(\ref{eq:lens}), (\ref{eq:focal_length}), and (\ref{eq:summary}),
by comparing their predictions to results of numerical
simulations. The latter were performed by solving the time-dependent
Schr{\"o}dinger equation, governing the evolution of the wave packet,
numerically, using the method of expanding the propagator in a series
of Chebyshev polynomials of the Hamiltonian. The reader is referred to
Refs.~\cite{TK84accurate, RKMF03Unified, DBS11Efficiency} for a
comprehensive description of the method and its implementations.

The concrete system that we consider is schematically illustrated in
Fig.~\ref{fig1} and described by the following set of
parameters. Hereinafter, we adopt atomic units, $\hbar = m = 1$. The
external Gaussian potential, Eq.~(\ref{eq:potential}), is
characterized by $l_1 = 0.1$ and $l_2 = 1$, and the potential strength
$V_0$ plays the role of a variable parameter, with values ranging
between 10 and 40. The initial state of the particle is given by the
wave function $\psi_0^{(1)} (q_1-Q; \rho_1) \, \psi_0^{(2)} (q_2;
\rho_2)$ with $Q = -0.8$, $v = 60$, and $\rho_1 = \rho_2 = -i (m v /
\hbar) (\sigma_0)^2$, where $\sigma_0 = 0.1$ quantifies the initial
position-space dispersion of the wave packet,
cf. Eq.~(\ref{eq:sigma}). Note that the kinetic energy of the
classical particle $E_0 = m v^2 / 2 = 1800$ is large compared to
the strength of the external potential.

\begin{figure}[ht!]
\includegraphics[width=3.3in]{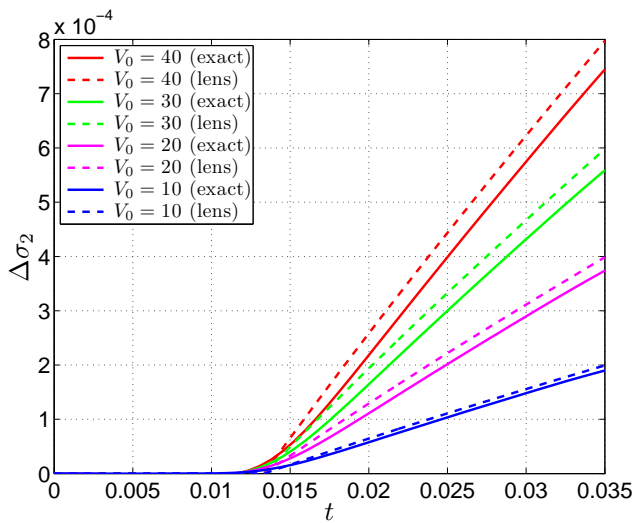}
\caption{(Color online) Spreading of the wave packet in the direction
  orthogonal to the direction of propagation. See text for details.}
\label{fig2}
\end{figure}

Our aim is to study the wave packet spreading in the direction
orthogonal to the propagation direction. This spreading is given by
the time-dependence of the dispersion along the $\ve_2$-axis,
\begin{equation}
  \sigma_2 = \sqrt{ 2 \int \ud \vq \, (q_2)^2 \left| \Psi_t (\vq)
    \right|^2 } \,.
\label{eq:sigma2}
\end{equation}
(Note that, due to the symmetry of $\Psi_t(\vq)$ under the reflection
$q_2 \rightarrow -q_2$, the expectation value of $q_2$ is zero, i.e.,
$\int \ud \vq \, q_2 | \Psi_t (\vq) |^2 = 0$.) For our choice of the
initial state, $\sigma_2 = \sigma_0$ at $t=0$, and $\sigma_2$
increases as a function of time. In the limit $V_0 = 0$, this increase
is determined by the free-particle spreading,
$\sigma_2^{\mathrm{free}} = \sigma(\rho_2 + vt)$, with the function
$\sigma(\rho)$ defined by Eq.~(\ref{eq:sigma}). A straightforward
calculation yields
\begin{equation}
  \sigma_2^{\mathrm{free}} = \sqrt{(\sigma_0)^2 + \left( \frac{\hbar
      t}{m \sigma_0} \right)^2} \,.
\label{eq:sigma2_free}
\end{equation}
In the case $V_0 \ne 0$, the external potential causes additional
spreading that can be quantified by
\begin{equation}
  \Delta \sigma_2 \equiv \sigma_2 - \sigma_2^{\mathrm{free}} \,.
\label{eq:Delta_sigma2}
\end{equation}

Figure~\ref{fig2} shows the dependence of $\Delta \sigma_2$ on time
$t$ for four different potential strengths, $V_0 =$ 10 (blue), 20
(magenta), 30 (green), and 40 (red). The solid curves represent the
results of the ``exact'' numerical solution of the time-dependent
Schr{\"o}dinger equation, while the dashed curves show the analytical
``lens'' approximation, namely $\sigma(\rho'_2) -
\sigma_2^{\mathrm{free}}$ with $\rho'_2$, calculated in accordance
with Eqs.~(\ref{eq:summary}), (\ref{eq:lens}), and
(\ref{eq:focal_length}). As expected, $\Delta \sigma_2 \simeq 0$ for
$t \lesssim |Q|/v \simeq 0.013$, corresponding to the time that it
takes for the classical particle to reach the potential
island. Figure~\ref{fig2} shows the theoretical predictions to be in a
reasonable agreement with the numerical results. It also confirms that
the agreement improves as the potential strength is decreased.

\section{Discussion and conclusions}

In this paper, we have constructed an approximate analytical solution
to the problem of a scattering of a localized quantum wave packet by a
weak Gaussian potential. Our solution is valid in the semiclassical
regime, in which the particle's de Broglie wavelength can be
considered short compared to all other length scales of the system. We
have shown that the quantum scattering process is closely analogous to
the phenomenon of focusing (or defocusing) of light rays by a thin
optical lens. In particular, the mathematical formula quantifying the
wave packet spreading, Eq.~(\ref{eq:lens}), is largely equivalent to
the thin lens formula of geometrical optics. The main difference
between the thin lens formula in optics and Eq.~(\ref{eq:lens}) is
that the former operates with true real-valued distances from the lens
to the object and to the image, while the ``distances'' $\rho_-$ and
$\rho_+$ entering Eq.~(\ref{eq:lens}) are intrinsically
complex-valued. It is only in the classical limit, $m v / \hbar
\rightarrow \infty$, that $\rho_-$ and $\rho_+$ become real-valued,
and that the optical thin lens formula is recovered.

It is instructive to further compare the ray optics picture with our
quantum mechanical result. The thin lens formula in optics, which is
the classical limit of Eq.~(\ref{eq:lens}), effectively describes the
deflection, or bending, of light rays (classical trajectories),
induced by the lens (external potential). However, only straight,
unbent trajectories have been used in our semiclassical derivation of
Eq.~(\ref{eq:lens}). This seeming paradox is resolved by the following
argument, originally presented in Ref.~\cite{BGAS95Chaotic} and for
readers' convenience reproduced in Appendix~\ref{app1}. The main
building block of the semiclassical propagator, Eq.~(\ref{eq:prop-2}),
is the Hamilton's principal function along the classical trajectory
connecting the initial and final points of the propagation. However,
in a sufficiently weak external potential, the value of the Hamilton's
principal function along the true (generally bent) classical
trajectory is very close to that along the corresponding straight
(unbent) trajectory, making the precise geometrical shape of the
trajectory unsubstantial.

The approach taken in this paper is conceptually similar to the one
used in Ref.~\cite{GD05Lyapunov} to analyze the spreading of quantum
wave packets in the Lorentz gas. The latter consists of a particle
moving in an array of fixed elastic scatterers, taken to be hard disks
(spheres) in two (three) spatial dimensions. However, it is important
to point out that in the Lorentz gas, unlike in the system addressed
in the present paper, one must take into account deflections of
classical trajectories in order to obtain the quantum-mechanical
equivalent of the circular (spherical) mirror formula.

\acknowledgments

The authors thank Tobias Kramer for useful
discussions. A.G. acknowledges the hospitality of the University of
Regensburg during a two-month visit where much of this work was done.
K.R. thanks the {\em Deutsche Forschungsgemeinschaft} for financial
support within Research Unit FOR 760.

\appendix

\section{Expansion of Hamilton's principal function}
\label{app1}

The following discussion is based on the argument that was, e.g.,
presented in Ref.~\cite{BGAS95Chaotic}.

Let us consider the Lagrangian 
\begin{equation}
  \Lag_\epsilon [\vr] (\tau) = \frac{m}{2} \left( \frac{\ud \vr}{\ud
    \tau}(\tau) \right)^2 - \epsilon V\big( \vr(\tau) \big)
\label{eq:app1.1}
\end{equation}
along a trajectory $\vr(\tau)$. Here $\epsilon \ll 1$ serves as a
dimensionless strength of the potential. The corresponding Hamilton's
principal function is given by
\begin{equation}
  S_\epsilon [\vr] (t) = \int\limits_0^t \ud \tau \, \Lag_\epsilon
  [\vr] (\tau) \,.
\label{app1.2}
\end{equation}

We now denote by $\vr_\epsilon (\tau)$ a trajectory that satisfies the
boundary conditions $\vr_\epsilon(0) = \vq'$ and $\vr_\epsilon(t) =
\vq$, and makes the action $S_\epsilon [\vr] (t)$ stationary, i.e.,
\begin{equation}
  \frac{\delta S_\epsilon}{\delta \vr} [\vr_\epsilon] (\tau) = {\bf 0}
  \,.
\label{app1.3}
\end{equation}
Then, for a trajectory $\vr_0 (\tau)$, that satisfies the same
boundary conditions, $\vr_0(0) = \vq'$ and $\vr_0(t) = \vq$, and that
is the stationary trajectory of $S_0 [\vr] (t)$, we have
\begin{align}
  &S_\epsilon [\vr_0] (t) = S_\epsilon [\vr_\epsilon] (t) + \int_0^t
  \ud \tau \, \frac{\delta S_\epsilon}{\delta \vr} [\vr_\epsilon]
  \cdot (\vr_0-\vr_\epsilon) \nonumber\\ &\quad + \frac{1}{2}
  \int\limits_0^t \ud \tau \, (\vr_0-\vr_\epsilon) \cdot
  \frac{\delta^2 S_\epsilon}{\delta \vr^2} [\vr_\epsilon] \;
  (\vr_0-\vr_\epsilon) + \ldots \label{app1.4} \,.
\end{align}
Substituting Eq.~(\ref{app1.3}) into (\ref{app1.4}) and taking into
account $\vr_\epsilon = \vr_0 + \mathcal{O}(\epsilon)$ we obtain
\begin{equation}
  S_\epsilon [\vr_\epsilon] (t) = S_\epsilon [\vr_0] (t) +
  \mathcal{O}(\epsilon^2) \,.
\label{app1.5}
\end{equation}



\end{document}